\documentstyle[aps,prl,epsf]{revtex}

\begin{document}
\draft
\wideabs{
\title{The quantum Hall plateau transition at order $1/N$}
\author{Joel E. Moore$^{1,2}$, A. Zee$^2$, and Jairo Sinova$^3$}
\address{$^1$Bell Labs, Lucent Technologies, 700 Mountain Avenue,
Murray Hill, NJ 07974 \\
$^2$ Institute for Theoretical Physics, University of California,
Santa Barbara, CA 93106 \\
$^3$ Department of Physics, The University of Texas at Austin,
Austin, TX 78712}
\date{\today}
\maketitle
\begin{abstract}
The localization behavior of noninteracting two-dimensional electrons
in a random potential and strong magnetic field is of fundamental
interest for the physics of the quantum Hall effect.  In order to
understand the emergence of power-law delocalization near the discrete
extended-state energies $E_n = \hbar \omega_c (n+\frac{1}{2})$, we
study a generalization of the disorder-averaged Liouvillian framework
for the lowest Landau level to $N$ flavors of electron densities
($N=1$ for the physical case).  We find analytically the large-$N$ limit
and $1/N$ corrections for all disorder strengths: at $N = \infty$ this
gives an estimate of the critical conductivity, and at order $1/N$ an
estimate of the localization exponent $\nu$.  The localization
properties of the analytically tractable $N \gg 1$ theory seem to
be continuously connected to those of the exact quantum Hall plateau
transition at $N = 1$.
\end{abstract}
}
The explanation by Laughlin~\cite{laughlin} of the integer quantum
Hall effect depends upon an understanding of the localization of
electrons by disorder in a strong magnetic field.  Without disorder,
the single-electron eigenstates fall into Landau levels at isolated energies
$E_n = \hbar \omega_c (n+\frac{1}{2})$, $n = 0,1,2,\ldots$, separated
by the cyclotron energy $\hbar \omega_c$.  The effect of a weak
disorder potential is to displace some weight from the
$\delta$-function peaks at $E_n$ into localized states at nearby
energies.  Extended states persist at energies $E_n$, and the
localization length $\xi(E)$ of states at energy $E$
diverges as $E \rightarrow E_n$ according to a power law $\xi(E)
\propto (E - E_n)^{-\nu}$.  Experimental results on disordered
samples~\cite{wei} are consistent with the value $\nu
\approx 2.35 \pm 0.05$ obtained from numerical
calculations~\cite{huckestein,chalker} on the lowest Landau level
(LLL).

Current belief is that the quantum Hall
plateau transition lies in an entirely different universality class
from the zero-field case, characterized by ``two-parameter scaling''
\cite{khmelnitskii,pruisken} and a topological term in the
$\sigma$-model description~\cite{pruisken}.  In particular there is
now an understanding of the minimal features required to obtain
numerically scaling behavior near the transition~\cite{chalker}, which
is in some respects similar to classical percolation, but with
different universal properties as a result of quantum tunneling and
interference~\cite{dhlee2}.  There is also some understanding of the
apparent insensitivity to interactions of some critical indices such
as $\nu$.~\cite{dhlee} However, relatively little progress has been made in
finding an analytically tractable description of the localization
properties near the critical energy.  The subject of this paper is a
generalization of the problem to multiple flavors of electron
densities, allowing an analytical approach to the transition.  Our
discussion is based on the Liouvillian approach introduced by
Sinova, Meden, and Girvin~\cite{sinova}, reviewed below.

The generalization of the disorder-averaged action to $N$ flavors of
electron densities gives a simple mean-field-like theory in the
large-$N$ limit.  Other large-$N$ approaches such as~\cite{hikami}
typically generalize the non-interacting problem before the disorder
average and do not obtain $\nu$.  At first order in the small
parameter $1/N$, we recover anomalous scaling of the localization
length, i.e., a value for the critical exponent $\nu$.  Thus we find
an analytically tractable quantum description whose physics seems to
connect smoothly in the parameter $N$ to the plateau transition ($N =
1$).  The control parameter $1/N$ allows a systematic expansion around
$N = \infty$, and the large-$N$ limit and $1/N$ corrections can be
found analytically for all disorder strengths.


The localization properties of electrons at energy $E$
are contained in the correlation function of the LLL-restricted
density operators ${\bar \rho}_{\bf q}$:
\begin{equation}
{\tilde \Pi}(q,t; E) = {- i \Theta(t) \over N_L \hbar \ell^2}
\big\langle\!\big\langle {\rm Tr}\,{\bar \rho}_{\bf q}(t)
{\bar \rho}_{-{\bf q}}(0) \delta(E-H) \big\rangle\!\big\rangle.
\label{eqone}
\end{equation}
Here $N_L$ is the number of states in the LLL ($N_L \rightarrow
\infty$ taken below), $\ell = \sqrt{\hbar c/eB}$, and $\langle \langle
\rangle \rangle$ indicates the quenched disorder average.

The Liouvillian approach uses the integral over $E$ of ${\tilde
\Pi}(q,t; E)$:
\begin{eqnarray}
{\tilde \Pi}(q,t) &\equiv& \int dE\, {\tilde \Pi}(q,t; E) \cr
&=& {-i \Theta(t) \over N_L \hbar \ell^2}
\langle\!\langle {\rm Tr}\,{\bar \rho}_{\bf q}(t)
{\bar \rho}_{-{\bf q}}(0) \rangle\!\rangle.
\label{piint}
\end{eqnarray}
The key to the approach is that ${\tilde \Pi(q,t)}$
still contains information about electron localization but is
more easily calculated than the fixed-energy quantity ${\tilde
\Pi(q,t; E)}$.  The disorder average for the Fourier transform
${\tilde \Pi}(q,\omega)$ was carried out numerically in \cite{sinova}
and shown to obey the form $\omega {\rm Im} {\tilde \Pi}(q,\omega) =
\omega^{1 / 2\nu} f(q^2/\omega)$, where $f$ is an unknown scaling
function.  At mean-field level \cite{sinova}, only diffusion is found:
our model gives a systematic expansion beyond this mean-field result.
A major goal of this paper is to obtain the prefactor
$\omega^{1/2\nu}$ by a $1/N$ expansion of ${\tilde \Pi}(q,\omega)$.


The scaling form for ${\tilde Pi}(q,\omega)$ comes about because at
large $R$, only states with energies satisfying
\begin{equation}
{|E - E_c| \over E_c} < \left({a_d \over R}\right)^{1/\nu}
\end{equation}
will have $\xi(E) > R$ and hence contribute to the correlation
function at long enough times.  Here $a_d$ is a nonuniversal length
set by the disorder potential.  Thus the integral over energy which
gives ${\tilde \Pi}(q,\omega)$ is only nonzero in a window of size
proportional to $q^{1/\nu}$.  For states delocalized on the scale $R =
1/q$, ${\tilde \Pi}(q,\omega; E)$ follows ordinary
diffusive scaling ($\omega {\rm Im} {\tilde \Pi}$ is a function of
$q^2/\omega$).  A prefactor $\omega^{1/2\nu}$ rather than $q^{1/\nu}$
gives the scaling function a simple form: for $q^2 \ll \omega$
\begin{equation}
\omega {\rm Im} {\tilde \Pi}(q,\omega) \approx D(\omega) (q^2/\omega),
\end{equation}
with the frequency-dependent diffusion constant
$D(\omega) = D_0 \omega^{1 / 2 \nu}.$  This form also applies
in the classical percolation limit studied by Gurarie and
one of us~\cite{gurarie}.

The LLL-projected density operators ${\bar \rho}_{\bf q}$ are related to
the operators $\tau_{\bf q}$ of the magnetic translation group
through~\cite{girvin}
${\bar \rho}_{\bf q} = e^{-\frac{1}{4} \ell^2 q^2} \tau_{\bf q}.$
The noninteracting LLL-projected Hamiltonian is $H = \sum_{\bf q} v(-q) {\bar \rho}_{\bf q}$,
with $v$ the Fourier-transformed random potential.
Then using the commutation relation for the operators $\tau_{\bf q}$
\begin{equation}
[\tau_{\bf q}, \tau_{\bf r}] = 2 i \sin\left(\frac{\ell^2}{2}
{\bf q} \wedge {\bf r} \right) \tau_{\bf q+r},
\end{equation}
we obtain the evolution equation for the magnetic translation operators:
\begin{equation}
{\dot \tau}_{\bf q} = - i \sum_{{\bf q}^\prime}
{\cal G}_{{\bf q} {\bf q}^\prime}
\tau_{{\bf q}^\prime},
\end{equation}
in terms of the Girvian
\begin{equation}
{\cal G}_{{\bf q} {\bf q}^\prime} = {2 i \over \hbar} v({\bf q} -
{\bf q}^\prime) e^{-\frac{1}{4} \ell^2 |{\bf q}^\prime - {\bf q}|^2}
\sin\left(\frac{\ell^2}{2} {\bf q}^\prime \wedge {\bf q}\right).
\end{equation}
Then ${\tilde \Pi}(q,\omega)$ is a one-body correlation
function of the Girvian:
\begin{equation}
{\tilde \Pi}(q,\omega) = {1 \over \hbar \ell^2} \Big\langle\!\Big\langle
\,\langle {\bf q} | (\omega - {\cal G})^{-1} | {\bf q} \rangle
\, \Big\rangle\!\Big\rangle.
\end{equation}
Here the states $|{\bf q}\rangle$ and operator ${\cal G}$
are defined through $\langle{\bf q}| {\cal G} |{\bf q}^\prime\rangle$
$= {\cal G}_{{\bf q}{\bf q}^\prime}.$

Now we take the continuum limit $N_L \rightarrow \infty$
and assume a white-noise disorder potential
$\langle\langle v({\bf q}) v(-{\bf q}) \rangle \rangle =
{2 \pi v^2 \over L^2} \delta({\bf q}+{\bf q}^\prime)$.
The physical frequency $\omega_0$ is replaced by the dimensionless
combination $\omega = {h \omega_0 \ell / v}$.  The physical propagator
is ${\tilde \Pi}(q,\omega_0) = {h \ell \over v} \Pi(q,\omega)$, with
the dimensionless propagator $\Pi(q,\omega) \rightarrow
\frac{1}{\omega}$ in the clean limit $\omega \rightarrow \infty$.
All momenta are dimensionless (scaled by magnetic length $\ell$)
in the following.

The result of disorder averaging is an interacting theory
which can be expressed through a functional integral over both bosonic
$\phi$ and
Grassmann $\psi$ variables~\cite{sinova}:
\begin{eqnarray}
\Pi(q,\omega) &=& -i \int D{\bar \phi}\, D\phi\, \int D{\bar\psi}\,D \psi\,
{\bar \phi}_q \phi_q e^{-F(\omega)}, \cr
F(\omega) &=& -i \omega \int\,d{\bf q}\,
({\bar \phi}_{\bf q} \phi_{\bf q} + {\bar \psi}_{\bf q} \psi_{\bf q})
+ \cr&&\int_{1,2,3,4} f(1,2,3,4)
\Big[
{\bar \phi}_{{\bf q}_1} {\bar \phi}_{{\bf q}_2}
{\phi}_{{\bf q}_3} {\phi}_{{\bf q}_4} + \cr
&&\quad\quad 2 {\bar \psi}_{{\bf q}_1} {\bar \phi}_{{\bf q}_2}
{\phi}_{{\bf q}_3} {\psi}_{{\bf q}_4} +
{\bar \psi}_{{\bf q}_1} {\bar \psi}_{{\bf q}_2}
{\psi}_{{\bf q}_3} {\psi}_{{\bf q}_4} \Big].
\end{eqnarray}
The effective interaction from disorder averaging is
\begin{eqnarray}
f(1,2,3,4) &=& {1 \over \pi} e^{-\frac{1}{2} |{\bf q}_1 - {\bf q}_4|^2}
\delta({\bf q_1} + {\bf q}_2 - {\bf q}_3 - {\bf q}_4) \cr
&&\times \sin\left(\frac{1}{2}{\bf q}_1 \wedge {\bf q}_4 \right)
\sin\left(\frac{1}{2}{\bf q}_2 \wedge {\bf q}_3 \right).
\end{eqnarray}
The effect of the additional Grassmann variables
(``supersymmetry'') is simply to eliminate processes with more than
one density line.  We develop a natural generalization of this
interacting problem to $N$ flavors of densities ($N=1$ is the original
problem).  The $N \rightarrow \infty$ limit gives an approximate
propagator which is similar to the self-consistent Born
approximation of~\cite{sinova}.  The utility of the large-$N$ approach
is that it gives a systematic expansion in the parameter $1/N$ around
the diffusive $N \rightarrow \infty$ result, while at the point $N=1$
the theory describes the exact localization properties of the
plateau transition.

The $N$-flavor generalization of the Lagrangian density
is (only the bosonic part is given, for compactness, and the
flavor indices $i,j$ run from 1 to $N$)
\begin{eqnarray}
{\cal L}_N &=& -i \omega 
{\bar \phi}^i_q \phi^i_q +
{f(1,2,3,4) \over N}
\Big[c_1 {\bar \phi}_{q_1}^i {\bar \phi}_{q_2}^j \phi_{q_3}^i \phi_{q_4}^j \cr
&&+ c_2 {\bar \phi}_{q_1}^i {\bar \phi}_{q_2}^i \phi_{q_3}^j \phi_{q_4}^j
+ c_3 {\bar \phi}_{q_1}^i {\bar \phi}_{q_2}^j \phi_{q_3}^j
\phi_{q_4}^i\Big].
\label{nlag}
\end{eqnarray}

The real coefficients $c_i$ in (\ref{nlag}) reproduce the correct $N
\rightarrow 1$ limit provided that $c_1 + c_2 + c_3 = 1$, so there is
a two-parameter family of generalizations.  The vertex with
coefficient $c_3$ in (\ref{nlag}) does not contribute as $N
\rightarrow \infty$ and does not seem to affect scaling qualitatively
at order $1/N$, so it is dropped for simplicity.~\cite{moore} We
specialize to $c_1 = c_2 = 1/2$ in what follows: the choice of these
coefficients equal is ``natural'' in that the classes of diagrams
selected by the two vertices have equal weight at
order $1/N$ with $N=1$, as they do in the full theory (i.e.,
all orders in $1/N$) at $N = 1$.  For generic $c_i$ the theory has
a $U(1|1) \times SO(N)$ symmetry, which at the point $c_3 = 1$ ($N$
decoupled systems) becomes $U(N|N)$.

\begin{figure}
\epsfxsize=3.0truein
\centerline{\epsffile{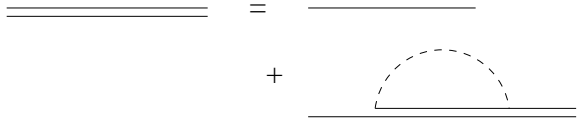}}
\caption{Diagrammatic representation of the large-$N$ propagator
equation (13).  The double line is the noncrossing
propagator.}
\label{figone}
\end{figure}

In the $N \rightarrow \infty$ limit, the diagrams with $k$ interaction lines
which contribute to $\langle {\bar \phi}^1_q \phi^1_q \rangle$ are the
diagrams where no interaction lines cross, which have the maximum $k$
free choices of flavor index (i.e., degeneracy $N^{k}$).  Only the
first interaction term of (\ref{nlag}) affects this limit.  The
noncrossing propagator sums these diagrams and satisfies an integral
equation depicted graphically in Fig.~\ref{figone}:
\begin{eqnarray}
\Pi_B(&q&,\omega) = {1 \over \omega} +
{\Pi_B(q,\omega) \over \pi \omega} \times \cr
&&\int\,d{\bf q}^\prime \, 
\sin^2\left(\frac12{{\bf q} \wedge {\bf q}^\prime}\right)
e^{-\frac12{|{\bf q} - {\bf q}^\prime|^2}}
\Pi_B(q^\prime,\omega).
\label{largen}
\end{eqnarray}
The notation $\Pi_B(q,\omega)$ is that of~\cite{sinova}.
In the limit $q,\omega \rightarrow 0$,
$q^2 \ll \omega$, the noncrossing result $\Pi_B(q,\omega)$ shows diffusive
behavior ($\omega \hbox{Im} \Pi = D_0 q^2 / \omega$)
without the prefactor $\omega^{1/2\nu}$.

The zeroth-order diffusion constant $D_0$ is a factor of $\sqrt{2}$
smaller here than in the self-consistent Born
approximation of~\cite{sinova} (the calculation there corresponds
to the choice $c_1 = 1$).  The resulting estimate of the critical
conductivity at the transition, obtained from $D_0$ and the exact
density of states~\cite{wegner} through the Einstein relation, is
$\sigma_{xx} \approx 0.614 \frac{e^2}{h}$ compared to the numerical
result $\sigma_{xx} = (0.54 \pm 0.04) \frac{e^2}{h}$~\cite{huo} and
$\sigma_{xx} = \frac{\sqrt{3} e^2}{4 h} \approx 0.433\frac{e^2}{h}$
of~\cite{gurarie}.  Although the large-$N$ estimates of $\sigma_{xx}$
from the Liouvillian are of the right order, it should be noted
that (unlike $\nu$) a universal $\sigma_{xx}$ has not yet been
found from $\Pi(q,\omega)$ obtained using numerical diagonalization.

\begin{figure}
\epsfxsize=3.0truein
\centerline{\epsffile{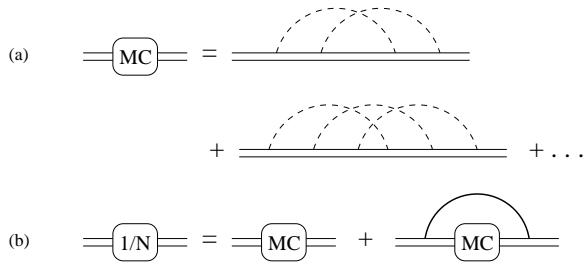}}
\caption{Diagrams contributing to $1/N$ propagator correction.  The
solid interaction line in (b) indicates one or more noncrossing interaction
lines.  Each diagram of order $1/N$ has one fewer free index choice
than the non-crossing diagrams of the same order.}
\label{figtwo}
\end{figure}

The noncrossing propagator has simple behavior at large $q$,
where the argument of the integrand in (\ref{largen}) is rapidly
oscillatory.  In this limit
\begin{equation}
\Pi_B(\infty,\omega) = {1 \over \omega} +
{\Pi_B(\infty,\omega)^2 \over \omega},
\end{equation}
or $\Pi_B(\infty,\omega) \approx - i + \frac{\omega}{2}$ for small
$\omega$.

\begin{figure}
\epsfxsize=2.5truein
\centerline{\epsffile{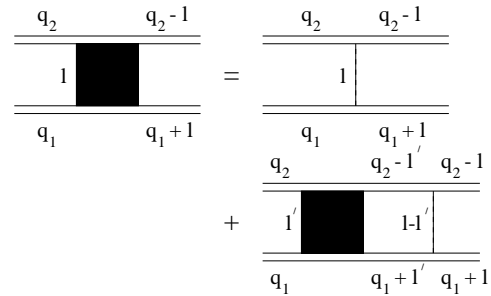}}
\caption{Schematic representation of sum of ladder diagrams.  The
solid block is defined as the sum of one or more rungs with the
specified total momentum transfer.}
\label{figthree}
\end{figure}

The first corrections to the noncrossing propagator have $k-1$ free
choices of index for a diagram of $k$ lines.  The corrections
consist of all maximally crossed diagrams of $2$ or more lines
(using the noncrossing propagator) (Fig.~\ref{figtwo}a),
and in addition possible ``rainbows''
over the maximally crossed portion (Fig.~\ref{figtwo}b).
The sum of all maximally crossed diagrams can be obtained
from the ladder sum represented in Fig.~\ref{figthree}
since maximally crossed diagrams are related to ladder diagrams after
cutting the center propagator line and pivoting.

The sum of all ladder diagrams with incoming momenta ${\bf q}_1, {\bf
q}_2$ and momentum transfer ${\bf l}$ as labeled in
Fig.~\ref{figthree}, denoted by $V({\bf q}_1,{\bf q}_2; {\bf l})$,
satisfies the integral equation
\begin{eqnarray}
V({\bf q}_1&,&{\bf q}_2; {\bf l}) = V_0({\bf q}_1,{\bf q}_2; {\bf l}) + \cr
&&\int\,{d{\bf l}^\prime \over \pi}
\big[V_0({\bf q}_1+{\bf l}^\prime,{\bf q}_2-{\bf l}^\prime;
{\bf l}-{\bf l}^\prime) \times \cr
&& \Pi_B(q_1+l^\prime,\omega)
\Pi_B(q_2-l^\prime,\omega)
V({\bf q}_1,{\bf q}_2;{\bf l}^\prime)\big]
\label{mcdiag}
\end{eqnarray}
where $V_0$ is the original interaction
\begin{equation}
V_0({\bf q}_1,{\bf q}_2; {\bf l}) \equiv
{e^{-\frac12{|{\bf l}|^2}} \over \pi}
\sin\left({{\bf q_1} \wedge {\bf l} \over 2}\right)
\sin\left({{\bf q_2} \wedge {\bf l} \over 2}\right).
\end{equation}
Note that $V$ has an implicit $\omega$ dependence.
The content of (\ref{mcdiag}), discussed below, is
that $V$ has a diffusion pole when ${\bf q}_2 \approx - {\bf q}_1$:
\begin{equation}
V({\bf q},{\bf q}^\prime; {\bf q}^\prime - {\bf q}) \propto
{1 \over i \omega + D_1 (q + q^\prime)^2}.
\label{diffpole}
\end{equation}

The physics of (\ref{diffpole}) is similar to that of the
weak-localization (WL) logarithmic singularity in two
dimensions~\cite{palee}, but with two major differences: the
disorder-generated effective interaction involves 4 bosonic density
operators and hence 8 rather than 4 fermionic operators, and
time-reversal symmetry is broken by the magnetic field.
The singularity results when the integral equation (\ref{mcdiag})
becomes nearly $V = V_0 + V$, i.e., when the integral operator on the
right-hand-side has an eigenvalue going to 1.  Consider a rung of a
long ladder diagram.  The intermediate propagator momenta on the $n$th rung
from an end of the ladder have magnitude proportional to $\sqrt{n}$
(momenta walk randomly in the plane) so most momenta in a long ladder
can be assumed large.  Each rung adds two
propagators $\Pi_B(\infty,\omega) \approx
-i + \omega/2$, and the interaction averages to $-1 + \frac{1}{8}(q +
q^\prime)^2.$ This gives the estimate $D_1 = \frac{1}{8}$ in
(\ref{diffpole}).

The above derivation assumes white-noise disorder in the Girvian,
although physical white noise disorder corresponds to disorder of
correlation length $\sim \ell$ in ${\cal G}$~\cite{sinova}.  Another
assumption is that the lack of an upper momentum cutoff in the
Liouvillian approach does not soften the singularity (\ref{diffpole})
to a logarithm.  The singularity is found numerically to persist
without these assumptions, suggesting that the high-momentum degrees
of freedom do not change the scaling behavior qualitatively, as
in~\cite{boldyrev}.  We call the limit with
point disorder and finite cutoff the WL limit as these assumptions are
standard in that context.  Now we calculate $\Pi_{MC}(q,\omega)$ and
obtain a singular log $\omega$ contribution to the diffusion constant.

The contribution of all maximally crossed diagrams to the propagator is
\begin{equation}
{\Pi_{MC}(q,\omega) \over
\Pi_B(q,\omega)^2} = \int\,d{\bf q}^\prime \Pi_B(q^\prime,\omega)
(V - V_0)({\bf q},{\bf q}^\prime; {\bf q}^\prime - {\bf q}).
\label{mcint}
\end{equation}
Here the subtraction of $V_0$ removes the first ladder diagram,
which has no crossings and is already in $\Pi_B$.
Integrating $q^\prime$ in (\ref{diffpole}) leads to a 
$\log \omega$ contribution in $\Pi_{MC}$.

The sum of maximally crossed diagrams (\ref{mcint}) already shows
nondiffusive scaling (an anomalous prefactor in $\omega$), which is
modified quantitatively but not qualitatively by adding rainbows over
the MC diagrams.  There is a logarithimic divergence in the sum of
maximally crossed diagrams, leading to a diffusion
constant
\begin{eqnarray}
D(\omega) &=& D_0 + {a \log \omega \over N} + O(N^{-2}) \cr
&\approx& D_0 \exp\big({a \log \omega \over N D_0}\big)
= D_0 \omega^{a / N D_0}
\end{eqnarray}
using the standard device 
of re-exponentiation in large-$N$ and RG theories to
estimate critical exponents.  Hence
\begin{equation}
{1 \over 2 \nu} = {a \over N D_0} + O(N^{-2}).
\end{equation}

\begin{figure}
\epsfxsize=3.0truein
\centerline{\epsffile{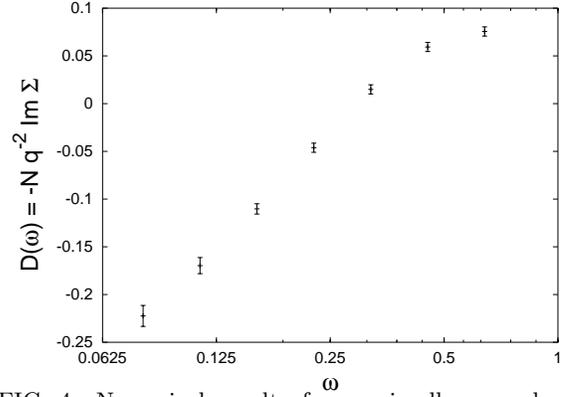}}
\caption{Numerical results for maximally-crossed
contribution to diffusion constant $D(\omega)$, on semilog scale.
Here $\Sigma = \omega - 1/\Pi.$}
\label{figfour}
\end{figure}

The coefficient $a$ is found from the numerical solution of
(\ref{mcdiag}) and (\ref{mcint}).  Fig.~\ref{figfour} shows sample
data from this calculation.  Numerically $D_0 = 0.682$, $a = 0.18 \pm
0.02$, and therefore $\nu = (1.89 \pm 0.1) N$.  In the WL limit the
addition of overloops (Fig. \ref{figtwo}b) is found analytically to
reduce $a$ and increase $\nu$ by a factor of 2.  Preliminary numerical
results~\cite{moore} are that the correction is in the same direction
for the full model.  Although this paper has focused on the LLL
plateau transition, the large-$N$ Liouvillian approach
may also be useful for other noninteracting quantum Hall transitions
with a discrete spectrum of extended states.

The authors wish to thank (J. E. M.)
M. Fogler, E. Fradkin, S. M. Girvin, D. H. Lee,
Z. Q. Wang, and X.-G. Wen, and (J. S.) V. Meden for useful discussions.
The authors acknowledge support from NSF grants
PHY-97-22022 and PHY-99-07949 (A. Z.), and the Welch Foundation (J. S.).


\begin{references}
\bibitem{laughlin}{R. B. Laughlin, Phys. Rev. B {\bf 23}, 5632 (1981).}
\bibitem{wei}{H. P. Wei, D. C. Tsui, M. A. Paalanen and A. M. M. Pruisken,
Phys. Rev. Lett. {\bf 61}, 1294 (1988); H. P. Wei, S. Y. Lin,
D. C. Tsui, and A. M. M. Pruisken, Phys. Rev. B {\bf 45}, 3926 (1992).}
\bibitem{huckestein}{B. Huckestein, Rev. Mod. Phys. {\bf 67}, 357 (1995).}
\bibitem{chalker}{J. T. Chalker and P. D. Coddington, J. Phys. C
{\bf 21} 2665 (1988).}
\bibitem{khmelnitskii}{D. E. Khmelnitskii, JETP {\bf 82}, 454 (1983).}
\bibitem{pruisken}{A. M. M. Pruisken, Nucl. Phys. B {\bf 235}, 227 (1984).}
\bibitem{dhlee2}{D. H. Lee, Z. Wang, and S. Kivelson,
Phys. Rev. Lett. {\bf 70}, 4130 (1993).}
\bibitem{dhlee}{D. H. Lee and Z. Q. Wang, Phys. Rev. Lett. {\bf 76},
4014 (1996).}
\bibitem{sinova}{J. Sinova, V. Meden, and S. M. Girvin,
Phys. Rev. B {\bf 62}, 2008 (2000).}
\bibitem{hikami}{S. Hikami, M. Shirai,
and F. Wegner, Nucl. Phys. B {\bf 408}, 415 (1993).}
\bibitem{gurarie}{V. Gurarie and A. Zee, {\tt cond-mat/0008163} (2000),
unpublished.  Their $G(k,t)$ disorder-averaged and multiplied by
a constant is equal to ${\tilde \Pi}(k,t)$.}
\bibitem{girvin}{S. M. Girvin and T. Jach, Phys. Rev. B
{\bf 29}, 5617 (1984).}
\bibitem{moore}{J. E. Moore, unpublished.}
\bibitem{wegner}{F. Wegner, Z. Phys. B {\bf 25}, 327 (1976).}
\bibitem{huo}{Y. Huo and R. N. Bhatt, Phys. Rev. Lett. {\bf 68},
1375 (1992).}
\bibitem{boldyrev}{S. Boldyrev and V. Gurarie,
{\tt cond-mat/0009203} (2000), unpublished.}
\bibitem{palee}{For a review see
P. A. Lee and T. V. Ramakrishnan, Rev. Mod. Phys.
{\bf 57}, 287 (1985).}

\end{references}
\end{document}